\documentstyle[12pt]{article}
\begin{document}
\title{
\begin{flushright}
{\small SMI-10-94 }
\end{flushright}
\vspace{2cm}
On Poisson-Lie structure on the external algebra of the classical Lie
groups }
\author{
G.E.Arutyunov \thanks{Steklov Mathematical Institute, Vavilov 42, GSP-1,
117966, Moscow,
Russia; arut@qft.mian.su}
 \\
and\\
P.B.Medvedev \thanks
{Institute of Theoretical and Experimental Physics,
117259 Moscow, Russia}
\thanks{Supported in part by RFFR under grant N93-011-147 }
}
\date {April 1994~}

\maketitle
\begin{abstract}
The general expression for the bicovariant bracket for odd generators
of the external algebra on a Poisson-Lie group is given.
It is shown that the graded Poisson-Lie structures
derived before for $GL(N)$ and $SL(N)$ are the special cases of this
bracket. The formula is the universal one and can be applied
to the case of any matrix Lie group.
\end{abstract}
\newpage

\section{Introduction}
The quantum bicovariant differential calculi first introduced by
Woronovicz \cite{Wor} is an object of primary importance in the
noncommutative differential geometry. Despite the existence of numerous
papers on this subject, up to now only the Quantum General Linear group was
equipped with the proper differential structure. As it can be seen from the
recent works \cite{IP,FP,MU,KU} the noncommutative differential geometry
reveals intriguing and unusual properties even for the case of
$SL_q(N)$.

In this note we continue to discuss the issue of Poisson-Lie
structure on external algebra of a Lie group. As it was realized
in our previous papers \cite{AM,AAM} the explicit description of
the graded Poisson-Lie brackets appears to be a powerful tool
in solving the problem of quantum bicovariant differential calculi \cite{Wor},
\cite{Car}-\cite{J}.
In this approach, in accordance with the general concept of Faddeev
\cite{FF}, objects in the theory of quantum groups appear as a result of
quantization of their classical counterparts.

In \cite{AM,AAM} our strategy  was the following. We consider the algebra
$Fun(G)$ of ordinary functions on a Lie group $G$ as a subalgebra of a
graded coalgebra ${\cal M}$ of external forms on $G$. The brackets on $
{\cal M}$ are defined as soon as one defines only three level of brackets:
of the order zero -- between a function and a function, of the first
order -- between a function and a one-form and of the second order --
between a one-form and a one-form. All the other brackets can be found by
using the graded Leibniz rule and linearity. The bracket of the
order zero on $Fun(G)$ is obviously fixed to be the Sklyanin bracket
\cite{Skl} and the others we define as a solution of a
bicovariance condition and the graded Jacobi identity.  This solution
for $G=GL(N),~SL(2)$
was obtained by straightforward and tedious calculations.

In this note we are going to elaborate an effective tool for solving this
problem, which can be applied to any classical matrix Lie group.
Especially, we confine ourselves by considering the graded brackets of the
second order. As we shall prove the bicovariance condition is strong
enough to fix the general form of these brackets:
\begin{equation}
\{\theta_1,\theta_2\}_{\mbox{c.b.}}=W^{0}_{~12}+ \left\{\theta_1[\theta_2,
r]\right\}
+\mbox{Tr}_{34}(W^2_{~1234}\theta_3\theta_4)+
\mbox{Tr}_{3456}(W^4_{~123456}\theta_3\theta_4\theta_5\theta_6)
\ldots,  \label{01}
\end{equation}
where $\theta$-s stand for the odd generators of ${\cal M}$.
The first term in (\ref{01}) is universal for any group ant it depends
only on a canonical $r$-matrix for the group in question. All the other
terms coming in eq.(\ref{01}) have the transparent geometric meaning as
the constant
Ad-invariant tensors with special symmetry properties. Thus, for a
given Lie group the question about the description of the brackets of the
second order reduces to the issue of the classification of special tensors
on this group.

We apply this technique to the case of $GL(N)$ and $SL(N)$ confining
ourselves by considering the terms in eq.(\ref{01}) of zero and second
orders in $\theta$-s. Under these assumptions it is not a problem to impose
the Jacobi identity that fixes in eq.(\ref{01}) some of free parameters.
The surprising fact is that the resulting Poisson brackets coincide (up to
$\theta$-independent terms) with the Poisson-Lie brackets
that we have obtained before.
In other words, the brackets of the second order are determined without any
reference to the brackets of the first order. The similar phenomena was
observed directly in the quantum case in \cite{IP}.

The paper is organized as follows. In Sec.2 the necessary definitions
and facts are collected. In Sec.3 we derive the formula (\ref{01})
for the bicovariant bracket of the second order. We apply this machinery
to the cases of $GL(N)$ and $SL(N)$ in Sec.4. Sec.5 is devoted to the
discussion.

\section   {Graded Poisson-Lie structure}
\setcounter{equation}{0}
The most transparent way to define a Poisson-Lie structure is to use
the notion of a coalgebra\footnote{We define a coalgebra as an algebra
supplied with a homomorphism $\Delta$ called the coproduct $\Delta~:
{\cal M}\rightarrow{\cal M}\otimes {\cal M}$ that is coassociative}.
We shall not list here all the basic
definitions, they can be found in the classical works \cite{Fad,Dr,Sem}, or
in our previous papers \cite{AM,AAM}. We shall pick up only some
necessary points below.

In the case of a matrix Lie group $G$
the appropriate coordinate system in $Fun(G)$
is given by the matrix elements of a matrix
$T=\parallel t_i^{~j}\parallel$ in the
fundamental representation of $G$.
In other words, $Fun(G)$ is defined to be an algebra generated by
the variables $t_i^{~j}$. Roughly speaking, functions on $G$ are
identified with formal power series in $t_i^{~j}$.
The Poisson-Lie brackets
in terms of
$t_i^{~j}$-s read \cite{Skl}:
\begin{equation}
\{ t_{i}^{~j},t_{k}^{~l}\}=r_{ik}^{~mn}t_{m}^{~j}t_{n}^{~l}
-t_{i}^{~m}t_{k}^{~n}r_{mn}^{~jl},
\label{cxxx}
\end{equation}
where $r_{ik}^{~mn}$ is the $r$-matrix which will be specified
later.
In the following it will be important that
the bracket (\ref{cxxx}) is degenerate and the function
$\det T$ lies in its center:
\begin{equation}
\{ t^{~j}_i ,\det T \}=0.
\label{xss}
\end{equation}
Fixing the value of $\det T$ equal to unity we obtain the Poisson-Lie
structure on $SL(N)$.

One fact from the ordinary differential geometry
will be necessary.
Let ${\cal G}^*$ be the
dual space of a Lie algebra ${\cal G}$. The cotangent bundle $T^{*}G$ on $G$ is
trivialized
by means of right (left) action of $G$ on
itself: $T^{*}G\approx G\times {\cal G}^*$. Let us
define the Maurer-Cartan
right-invariant form $\theta_{g}$ on $G$ which takes
value in ${\cal G}$:
$$
\theta_{g}(X_g)=X_e,
$$
where $X_g$ is a right-invariant vector field corresponding
to the element $X_e\in {\cal G}$.
With respect to the left action $g\rightarrow g_1g$ the form
$\theta_{g}$ transforms
as follows:
\begin{equation}
\theta_{g}\rightarrow \theta_{g_1g}=g_1\theta_{g}g_1^{-1}+dg_1g_1^{-1}.
\label{cl}
\end{equation}
(This equation is well-known in physical literature as
the gauge transformation law.)
One can treat the right hand side of eq.(\ref{cl}) as the differential
form on $G\times G$. With the identification
$T^*(G\times G)\approx T^*G\otimes T^*G$
eq.(\ref{cl}) can be
written as
\begin{equation}
\theta_g\rightarrow \theta_{g_1g}=(g_1\otimes I)(I\otimes
\theta_{g})(g_{1}^{-1}\otimes I)
+\theta_{g_{1}}\otimes I.
\label{cll}
\end{equation}

Now we can introduce our basic object
${\cal M}$.  To describe the external algebra of right-invariant forms we
add to the system of coordinates $t_{i}^{~j}$ new anticommuting variables
$\theta _{i}^{~j}$. To specialize the group we shall put the proper
constraints on $t$-s and $\theta$-s and now we shall treat them as
independent variables as it is for the case of $GL(N)$.
Hence, by
definition ${\cal M}$ is a free associative algebra generated by
$t_{i}^{~j},\theta _{i}^{~j},t$ modulo the relations:
$$ t_{i}^{~j}
t_{k}^{~l} =t_{k}^{~l}
t_{i}^{~j},~~tt_{i}^{~j}=t_{i}^{~j}t,~~t\det T=I,
$$
$$
t_{i}^{~j}\theta _{k}^{~l}=\theta_{k}^{~l}t_{i}^{~j},~~
t\theta _{i}^{~j}=\theta _{i}^{~j}t,~~
\theta _{i}^{~j}\theta _{k}^{~l}=-\theta_{k}^{~l}\theta
_{i}^{~j}.
$$
The algebra ${\cal M}$ has a natural
grading with $\deg{(t_{i}^{~j})}=0$ and
$\deg{(\theta_{i}^{~j})}=1$.

Let us define the  multiplication law in ${\cal
M}\otimes {\cal M}$ as follows:
\begin{equation} (a\otimes b)(c\otimes d)
=(-1)^{\deg{b}\deg{c}}(ac\otimes bd)
\label{br}
\end{equation}
for any
$a,b,c,d\in {\cal M}$.
It is well-known that $Fun(G)$ has the Hopf algebra structure with the
coproduct $\Delta$, the counit $\varepsilon$ and the antipode $S$:
\begin{equation}
\Delta
t_{i}^{~j}=t_{i}^{~k}\otimes t_{k}^{~l},~~ \varepsilon (t_{i}^{~j})=\delta
_{i}^{~j},~~S(t_{i}^{~j})=t\hat{t}_{j}^{~i}
\label {t}
\end{equation}
\begin{equation}
\Delta(t)=t\otimes t,~~\varepsilon (t)=1,~~S(t)=\det{T},
\label {tkl}
\end{equation}
where in the usual notation one has
$\parallel t_i^{~j}\parallel^{-1}=t\hat{t}_j^{~i}$.
One can endow ${\cal M}$  with two graded coalgebra structures
defining
the action of $\Delta_{L,R}$ on the generators
$t_{i}^{~j},\theta _{i}^{~j}, t$ as follows:
\begin{equation}
\Delta_R \theta _{i}^{~j}=\theta _{i}^{~j}\otimes I
\label {tt}
\end{equation}
\begin{equation}
\Delta_L \theta _{i}^{~j}= t_{i}^{~k}S(t_{p}^{~j})\otimes \theta _{k}^{~p},
\label{tt1}
\end{equation}
\begin{equation}
\Delta_R t_{i}^{~j}=\Delta_L t_{i}^{~j}= \Delta t_{i}^{~j},
\label {ttt}
\end{equation}
{\em i.e.} on the generators $ t_{i}^{~j}$ the action of $\Delta_{R,L}$
coincides with the action of $\Delta$.
To an arbitrary element of ${\cal M}$ the actions of $\Delta_{L,R} $
are extended as to be homomorphisms.

Our main goal is to equip ${\cal M}$ with a Poisson
structure consistent with the coproduct on ${\cal M}$. Precisely, it means
the following. We introduce  a bilinear operation $\{~,\}$: ${\cal M}\otimes
{\cal M}\rightarrow {\cal M}$ called  brackets.
The algebra $\cal M$ has odd and
even generators, so it is natural to require for this bracket the
fulfillment of the super Jacobi identity:
\begin{equation}
(-1)^{\deg{a}\deg{c}}\{\{ a,b\}, c \}+(-1)^{\deg{b}\deg{c}} \{\{ c,a\}, b
\}+ (-1)^{\deg{a}\deg{b}}\{\{ b,c\}, a \}=0,
\label{ya}
\end{equation}
the graded Leibniz rule
\begin{equation}
\{a\cdot b, c \}=a\{b, c \}+(-1)^{\deg{b}\deg{c}}\{a, c \}b,
\label{ya1}
\end{equation}
and the graded symmetry property:
\begin{equation}
\{a, b \}=(-1)^{\deg{a}\deg{b}+1}\{b, a \} ,~~~ \deg{\{a, b
\}}=(\deg{a}+\deg{b})\bmod 2.
\label{ya2}
\end{equation}

Let us require now that
our algebra $\cal M$ supplied with the Poisson structure would be a
Poisson-Lie algebra, i.e. the Poisson brackets would satisfy
\begin{equation}
\Delta_{R,L} \{ a,b\}_{{\cal M}}=\{\Delta_{R,L} (a),\Delta_{R,L} (b)\}_
{{\cal M\otimes}{\cal M}},
\label{cx}
\end{equation}
where the bracket on ${\cal M\otimes}{\cal M}$ is defined as
\begin{equation}
\{ a\otimes b,c\otimes d\}_{{\cal M}\otimes {\cal M}}=
(-1)^{\deg{b}\deg{c}}\{ a,c\}_{\cal M}\otimes bd+
(-1)^{\deg{b}\deg{c}}ac\otimes \{ b,d\}
\label{xx}
\end{equation}
for any elements $a,b,c,d\in {\cal M}$.
In what follows we will call
the algebra $\cal M$ equipped with the brackets defined above as the
Poisson-Lie superalgebra.

A comment is in order. It would be more natural to define the
coproduct $\Delta$ on right-invariant forms $\theta$ as
$\Delta =\Delta_L +\Delta_R$, in that case
the coproduct law $\Delta$ would
mimic the transformation law for the right invariant forms on a Lie group
(eqs.(\ref{cl}), (\ref{cll})). However, as we have found for the
$SL(2)$ there is no solutions of eq. (\ref{cx}) with the total
$\Delta$, hence we relax (\ref{cx}) by replacing the total $\Delta$
by two separate coproducts $\Delta_R$ and $\Delta_L$. This
choice of coalgebra structure strictly corresponds to the notion of
{\em bicovariant} differential calculi of Woronowicz \cite{Wor}.

To define the
brackets on $\cal M$ it is enough to define them on the set of
generators $t_{i}^{~j},\theta _{i}^{~j}$
and then to extend by the Leibniz rule
to the whole algebra.

The linear space $\cal N$ spanned by the generators $t_i^{~j}$ and
$t$ forms the Hopf subalgebra $Fun(G)$. Hence, we equip $\cal N$
with the Sclyanin bracket (\ref{cxxx}) described above:
\begin{equation}\label{R}
\{T_1 ,T_2\}=[r,T_1 T_2 ].
\end{equation}
Here we use the standard tensor notation: $T_1 =T\otimes I,~T_2 =I
\otimes T$.

\section{Bicovariant bracket of the second order}
\setcounter{equation}{0}
In this section we will derive the general expression
for the bicovariant bracket between two odd generators. We will
consider the situation when this bracket is required to be $Z_2$ graded
one rather then quadratic.

Note once more that we are going to solve the bicovariance condition
eq. (\ref{cx}) only. We shall denote this bi{\bf c}ovariant {\bf b}racket
as $ \{ \}_{c.b.}$. To obtain the actual Poisson bracket one needs also to
impose the Jacobi identity. We shall come back to this issue when
considering the concrete Poisson-Lie groups. Up to now we do not know the
general procedure to solve the Jacobi identity for the graded bracket
without specifying a group.

We start with imposing the $\Delta_R$ covariance. The eq. (\ref{cx}) reads:
\begin{equation}
\Delta_R\{ \theta_1 ,\theta_2\}_{\mbox{c.b.}}=\{\Delta_R (\theta_1 ),
\Delta_R (\theta_2 )\}_{\mbox{c.b.} {\cal M}\otimes{\cal M}}
\end{equation}
Here we use for $\theta$-s the same tensor notation as for $T$-s. By using the
explicit form for $\Delta_R$ and the definition of the bracket on
$\cal{M}\otimes\cal{M}$ we get
\begin{equation}\label{}
\Delta_R\{ \theta_1 ,\theta_2\}_{\mbox{c.b.}}=\{\theta_1 ,
\theta_2 \}_{\mbox{c.b.}}\otimes I
\end{equation}
that means that the $\{ \theta_1 ,\theta_2\}_{\mbox{c.b.}}$ is nothing but the
right-invariant form as the $\theta$-s are. We will use this key feature
below.

Now let us turn to the $\Delta_L$-covariance:
\begin{equation}
\label{YY}
\Delta_L\{ \theta_1 ,\theta_2\}_{\mbox{c.b.}}=
\{\Delta_L (\theta_1 ),
\Delta_L (\theta_2 )\}_{\mbox{c.b.} \cal{M}\otimes\cal{M}} .
\end{equation}
In the following it will be useful to adopt the convention: we shall omit
the sign of a tensor product in eq. (\ref{tt1}) writing it simply as
\begin{equation}\label{d}
\Delta_L \theta =T\theta S(T) .
\end{equation}
Clearly one should have in mind that $\theta$ and $T, S(T)$ belong to
different factors of the tensor product in $\cal{M}\otimes\cal{M}$.
Let us find the r.h.s. of eq. (\ref{YY}) :
\begin{equation}
\{\Delta_L
\theta_1,\Delta_L \theta_2\}_{\mbox{c.b.}}=
\{T_1\theta_1S(T_1),T_2\theta_2S(T_2)\}_{\mbox{c.b.}}.
\label{WW}
\end{equation}
As it is seen from this expression to get the answer we need to calculate
the ordinary brackets of the type $\{TS(T),TS(T)\}$. By using the
Leibniz rule and making
the proper arrangement of the matrix multipliers we have
$$
\{\Delta_L (\theta_1 ),
\Delta_L (\theta_2 )\}_{\mbox{c.b.} \cal{M}\otimes\cal{M}}=
\{T_1,T_2\}\theta_1\theta_2S(T)_1S(T)_2+T_1\theta_1\{S(T)_1,T_2\}\theta_2S(T)_2-
$$
$$
T_2\theta_2\{T_1,S(T)_2\}T_1S(T)_1-T_1T_2\theta_1\theta_2\{S(T)_1,S(T)_2\}+
T_1T_2\{\theta_1,\theta_2\}_{\mbox{c.b.}}S(T)_1S(T)_2.
$$
All the even brackets here are defined by the Sklyanin bracket
and look like
\begin{equation}
\{T_1,T_2\}=[r,T_1T_2],~~~
\{T_1,S(T)_2\}=-S(T)_2[r,T_1T_2]S(T)_2,
\label{XZX}
\end{equation}
$$
\{S(T)_1,S(T)_2\}=S(T)_1S(T)_2[r,T_1T_2]S(T)_1S(T)_2.
$$
Inserting eq. (\ref{XZX}) into the previous formula one gets
\begin{equation}
\{\Delta_L \theta_1,\Delta_L \theta_2\}_{\mbox{c.b.}}=
\label{uu}
\end{equation}
$$
T_1T_2\left(\{\theta_1,\theta_2\}_{\mbox{c.b.}}-
\left\{\theta_1[\theta_2, r]\right\} \right)S(T_1)S(T_2)+
\{T_1\theta_1S(T)_1,[T_2\theta_2S(T)_2,r]\},
$$
where $[,]$ stands for the commutator in a Lie algebra and $\{,\}$ -- for
the anticommutator.

Now let us note that by using our convention eq. (\ref{d})
one can write the last term in eq. (\ref{uu}) as
$$
\left(\left\{\Delta_L\theta_1,[\Delta_L\theta_2, r]\right\}\right)=
\Delta_L\left(\left\{\theta_1[\theta_2, r]\right\}\right)
$$
since $\Delta_L$ is an algebra homomorphism.

Thus, the left-covariance of the bracket $
\{\Delta_L \theta_1,\Delta_L \theta_2\}_{\mbox{c.b.}}=
\Delta_L\{\theta_1,\theta_2\}_{\mbox{c.b.}}
$
implies the fulfillment of the following relation
\begin{equation}
T_1T_2\left(\{\theta_1,\theta_2\}_{\mbox{c.b.}}-
\left\{\theta_1[\theta_2, r]\right\} \right)S(T_1)S(T_2)=
\Delta_L\left(\{\theta_1,\theta_2\}_{\mbox{c.b.}}-
\left\{\theta_1[\theta_2, r]\right\} \right).
\label{kk}
\end{equation}
In terms of the new tensor $\Theta$:
$$
\Theta(\theta)_{12}=\{\theta_1,\theta_2\}_{\mbox{c.b.}}-
\left\{\theta_1[\theta_2, r]\right\}
$$
the last equation takes the form
\begin{equation}
T_1T_2\Theta_{12}S(T_1)S(T_2)=\Delta_L \Theta_{12}
\end{equation}
or
\begin{equation}
\Theta_{12}=S(T_1)S(T_2)\Delta_L \Theta_{12}T_1T_2.
\label{10}
\end{equation}

Any tensor $\Theta$ can be written as the polynomial in $\theta$-s
with coefficients in $Fun(G)$. The grading requirement means that
$\Theta_{12}$  is an even element, {\em i.e.}
it has the form
\begin{equation}
\Theta_{ik}^{~jl}=W_{~~ik}^{0~jl}+W_{~~ik~i_1i_2}^{2~jl~j_1j_2}\theta_{j_1}^{~i_1}
\theta_{j_2}^{~i_2}+W_{~~ik~i_1i_2i_3i_4}^{4~jl~j_1j_2j_3j_4}
\theta_{j_1}^{~i_1}\theta_{j_2}^{~i_2}\theta_{j_3}^{~i_3}\theta_{j_4}^{~i_4}+
\ldots.
\label{11}
\end{equation}
Moreover, the right-covariance of the bracket forces all the tensors
$W^{2k}$ to be constant ($T$-independent) tensors. Hence, when acting on
$\Theta$ the coproduct $\Delta_L$ actually acts on $\theta$-s only.
Writing down this action explicitly we see that the equation (\ref{10}) is
nothing but the condition of the Ad-invariance of any tensor $W^{2k}$:
\begin{equation}
W^{2k}=(\underbrace{\mbox{Ad}_g\otimes \ldots
\otimes\mbox{Ad}_g}_{2k+2})W^{2k}.
\label{12}
\end{equation}

Now  let us note that the Sklyanin bracket depends only on the
antysimmetric part of the $r$-matrix. We take
r-matrix $r\in {\cal G}\wedge {\cal G}$ satisfying
the MCYBE.
Then the bracket $\left\{\theta_1[\theta_2, r]\right\} $ is
symmetric.
It means that $W^{2k}\in S^2{\cal G}\otimes
\wedge^{2k}{\cal G}$, where $S^2{\cal G}$ stands for the symmetric
part of the tensor product ${\cal G}\otimes {\cal G}$ and $\wedge^2{\cal
G}$ for the antysimmetric part respectively.  Thus the general form of the
bracket
between two odd generators is
\begin{equation}
\{\theta_1,\theta_2\}_{\mbox{c.b.}}=W^{0}_{~12}+ \left\{\theta_1[\theta_2,
r]\right\}
+\mbox{Tr}_{34}(W^2_{~1234}\theta_3\theta_4)+
\mbox{Tr}_{3456}(W^4_{~123456}\theta_3\theta_4\theta_5\theta_6)
\ldots.  \label{13}
\end{equation}

\section{Poisson bracket of the second order for $GL(N)$ and $SL(N)$}
\setcounter{equation}{0}
In this section we are going to show that when specifying a Poisson-Lie
group in the known cases the formula (\ref{13}) reproduces the correct
answers for the Poisson brackets on the external algebra of the
corresponding group. In particular, we will be interested here in the
Poisson-Lie groups $GL(N)$ and $SL(N)$. We also confine ourselves
by considering the {\em quadratic} brackets. It means that we
choose all the tensors $W^{2k}$ equal to zero when $k>1$. Under this
assumption the bracket (\ref{13}) takes the form
\begin{equation}
\{\theta_1,\theta_2\}_{\mbox{c.b.}}=W^{0}_{~12}+ \left\{\theta_1[\theta_2,
r]\right\}
+\mbox{Tr}_{34}(W^2_{~1234}\theta_3\theta_4).
\label{d13}
\end{equation}

\subsection{$GL(N)$ case}
Let us consider the Lie group $GL(N)$ and its Lie algebra ${\cal G}=gl(N)$.
Let $\{e_{\mu}\}$ be an orthonormal basis in ${\cal G}$. The Poisson-Lie
structure on the function algebra of $GL(N)$ is defined by
means of the Sklyanin bracket (\ref{R}) with the $r$-matrix being
the trivial lift of the canonical $sl(N)$ $r$-matrix. As mentioned
above we choose $r\in {\cal G}\wedge{\cal G}$ being the solution
of the Modified Classical Yang-Baxter Equation (MCYBE):
\begin{equation}
[r_{12},r_{13}]+[r_{12},r_{23}]+[r_{13},r_{23}]=\alpha[t_{13},t_{23}].
\label{MY}
\end{equation}
Here $[t_{13},t_{23}]$ is the Ad-invariant tensor in
${\cal G}\wedge{\cal G}\wedge{\cal G}$ defined via the canonical
element $t=e_{\mu}\otimes e_{\mu}\in {\cal G}\otimes {\cal G}$.
The constant $\alpha$ in the r.h.s. of eq.(\ref{MY}) defines
the normalization of the chosen $r$-matrix.

For the bracket (\ref{d13}) to be treated as the $GL(N)$-bicovariant
bracket one needs to specify the $\mbox{Ad}_{GL(N)}$-invariant tensors
$W^0\in S^2{\cal G}$ and $W^{2}\in S^2{\cal G}\otimes \wedge^2{\cal G}$.
For this purpose let us choose a basis in
$GL(N)$ consisting of the matrix unities $e_{i}^{~j}$:
$(e_{i}^{~j})_{k}^{~l}=\delta_{i}^{~l}\delta_{k}^{~j},~i,j=1,\ldots, N$.
The entries of the basis
elements coincide with the matrix elements of
the permutation operator $P$, {\em i.e.} $P_{ik}^{~jl}=
\delta_{i}^{~l}\delta_{k}^{~j}$. It is easy to see that $P$ is
Ad-invariant element in $S^2{\cal G}$. Except the permutation operator
there is only one element in $S^2{\cal G}$,namely the unity I, which is
Ad-invariant. Thus, in our basis one can wright
\begin{equation}
W^0_{12}(\beta_1,\beta_2)=\beta_1I+\beta_2P_{12},
\end{equation}
where $\beta_1,\beta_2$ are arbitrary numbers.

To proceed with the construction of $W^2$ let us note that two
Ad-invariant elements $I$ and $P$ in ${\cal G}\otimes {\cal G}$ can be
written in our basis via the matrix trace as
$$
I_{ik}^{~jl}=\delta_{i}^{~j}\delta_{k}^{~l}=
\mbox{Tr}(e_{i}^{~j})\mbox{Tr}(e_{k}^{~l})
$$
and
$$
P_{ik}^{~jl}=\delta_{i}^{~l}\delta_{k}^{~j}=
\mbox{Tr}(e_{i}^{~j}e_{k}^{~l}).
$$
In this form their Ad-invariance is obvious. Moreover, these formulae
imply the straightforward generalization to the higher invariants
of tensor products of ${\cal G}$. Clearly,
all invariants in ${\cal G}^{\otimes 4}$ should be of the types
\begin{equation}
\begin{array}{l}
L_1=\mbox{Tr}e\mbox{Tr}(e)\mbox{Tr}(e)\mbox{Tr}(e),\\
L_2=\mbox{Tr}(ee)\mbox{Tr}(e)\mbox{Tr}(e),\\
L_3=\mbox{Tr}(e)\mbox{Tr}(eee),\\
L_4=\mbox{Tr}(eeee).
\end{array}
\label{ii}
\end{equation}
We omitted here the indexes labelling the elements of the basis.
In such a way we have constructed all invariant tensors in
${\cal G}^{\otimes 4}$. Now the question is how to find the invariants
in the space $S^2{\cal G}\otimes \wedge^2{\cal G}$. Obviously, it can be
done by
symmetrization of the invariant tensors (\ref{ii}) with respect
to the first two spaces in ${\cal G}^{\otimes 4}$ and
antysimmetrization  with respect to the two others.
Applying successively this procedure one can
easily see that the first two invariants in (\ref{ii}) give zero
since there is no invariant tensors in $\wedge^2{\cal G}$.
As to the $L_3$ and $L_4$ the explicit computation
gives
\begin{equation}
(L_3)_{ikms}^{~jlnp}=
\delta_{i}^{~j}\delta_{k}^{~n}\delta_{s}^{~l}\delta_{m}^{~p}+
\delta_{i}^{~n}\delta_{s}^{~j}\delta_{k}^{~l}\delta_{m}^{~p}
-\delta_{i}^{~j}\delta_{k}^{~p}\delta_{m}^{~l}\delta_{s}^{~n}
-\delta_{i}^{~p}\delta_{k}^{~l}\delta_{m}^{~j}\delta_{s}^{~n},
\label{18}
\end{equation} and
\begin{equation}
(L_{4})_{ikms}^{~jlnp}=
\delta_{i}^{~l}\delta_{s}^{~j}\delta_{k}^{~n}\delta_{m}^{~p}+
\delta_{i}^{~n}\delta_{k}^{~j}\delta_{s}^{~l}\delta_{m}^{~p}
-\delta_{i}^{~l}\delta_{m}^{~j}\delta_{k}^{~p}\delta_{s}^{~n}
-\delta_{i}^{~p}\delta_{k}^{~j}\delta_{m}^{~l}\delta_{s}^{~n}.
\label{19}
\end{equation}

Combining  $L_3$ and $L_4$ with arbitrary numerical coefficients
$\beta_3$ and $\beta_4$ we obtain the general expression for the
tensor $W^2$:
$$
W^2=\beta_3L_3+\beta_4L_4.
$$

Now having at hand the
explicit form of $W^0$ and $W^2$ we substitute them in (\ref{d13})
and obtain the four-parameter family of quadratic covariant brackets
on the external algebra of $GL(N)$:
\begin{equation}
\{\theta_1,\theta_2\}_{\mbox{c.b.}}=\beta_1I+\beta_2P_{12}+
\left\{\theta_1[\theta_2, r]\right\}
+\beta_3(\theta_1\theta_1+\theta_2\theta_2)
+\beta_4(\theta_1P\theta_2+\theta_2P\theta_1).
\label{20}
\end{equation}
Now what we are going to do is to check the Jacobi identity:
\begin{equation}
J=\{\{\theta_1,\theta_2\}_{\mbox{c.b.}},\theta_3\}_{\mbox{c.b.}}+c.p.=0,
\label{Ji}
\end{equation}
where c.p. stands for the cyclic permutations of the indexes 1,2,3.
The direct calculation leads to the following answer for $J$
\begin{equation}
J=[\theta_1,\{\theta_2,[\theta_3,C(r)]\}]-\beta_{3}^2
[\theta_1,\{\theta_2,[\theta_3,\Omega]\}]=0,
\label{Ji1}
\end{equation}
where $C(r)$ is the l.h.s. of the MCYBE (\ref{MY}) and
$\Omega=[P_{13},P_{23}]$ is the Ad-invariant tensor standing in the
r.h.s. of MCYBE.
Thus, substituting $C(r)$ from eq.(\ref{MY}) into eq.(\ref{Ji1})
we arrive to
\begin{equation}
J=(\alpha-\beta_4^2)[\theta_1,\{\theta_2,[\theta_3,\Omega]\}].
\label{Ji2}
\end{equation}
Since the tensor in the l.h.s. of the last equation is not equal to zero
the Jacobi identity is satisfied if and only if $\beta_4=\pm \sqrt{\alpha}$
whereas $\beta_1,\beta_2,\beta_3$ remain to be arbitrary.
Therefore, the graded bicovariant brackets given by
\begin{equation}
\{\theta_1,\theta_2\}_{\mbox{c.b.}}=
(\beta_1I+\beta_2P_{12})
+\{\theta_1[\theta_2,r]\}+\beta_3(\theta_1\theta_1+\theta_2\theta_2)
\pm \sqrt{\alpha}(\theta_1P\theta_2+\theta_2P\theta_1)
\label{2gg}
\end{equation}
defines the genuine graded Poisson-Lie structure  in quadratic
sector of $\theta$ generators.

In \cite{AM} using the full coproduct law
$\Delta=\Delta_R+\Delta_L$ \footnote{In this case ${\cal M}$ can be
equipped with the graded Hopf algebra structure.} the following
one-parameter family of the graded Poisson-Lie brackets between
two odd generators was derived:
\begin{equation}
\{\theta_1,\theta_2\}_{\mbox{c.b.}}=\beta(\theta_1\theta_1+\theta_2\theta_2)+
r_{+}^{12}\theta_1\theta_2+\theta_1\theta_2r_{+}^{12}+
\theta_2r_{\pm}^{12}\theta_1-\theta_1r_{\mp}^{12}\theta_2.
\label{3gg}
\end{equation}

Now comparing (\ref{2gg}) with (\ref{3gg}) one can see that the
requirement for the bracket $\{\theta,\theta\}$ to be $\Delta_{R,L}$
covariant instead of being $\Delta$-covariant leads to
the appearance in (\ref{2gg}) the additional terms of another degree.
However, in the pure quadratic sector (when we put
in (\ref{2gg}) $\beta_1=\beta_2=0$) by using the definition of
$r_{\pm}=r\pm \sqrt{\alpha}P$  we conclude that
equations (\ref{2gg}) and (\ref{3gg}) coincide.

\subsection{$SL(N)$ case}
Here we will study the possibility of reducing the general expression
(\ref{d13}) to the bicovariant bracket on the external algebra of $SL(N)$.
In principle there is no difference in the procedure of
description of invariant elements for  $GL(N)$ and $SL(N)$ cases.
To begin with let us choose the following set of matrices in the fundamental
representation of ${\cal G}=sl(N)$:
$(E_{i}^{~j})_{k}^{~l}=\delta_{i}^{~l}\delta_{k}^{~j}-\frac{1}{N}
\delta_{i}^{~j}\delta_{k}^{~l}, i,j=1,\ldots,N.$
Note that the matrix elements of the matrices from this set can be written
as follows
$$
(E_{i}^{~j})_{k}^{~l}=P_{ik}^{~jl}-\frac{1}{N}I_{ik}^{~jl}.
$$
Clearly, the matrices $E_{i}^{~j}$ do not constitute the basis in
${\cal G}$ because they are linear dependent $E_{i}^{~i}=0$. Nevertheless,
we can use them to construct invariant tensors of $SL(N)$ just in the same
way as the basis of matrix unities was used for the $GL(N)$ case.
There is only one invariant tensor in $S^{\cal G}$:
$$\mbox{Tr}(E_{i}^{~j}E_{k}^{~l})-\frac{1}{N}\mbox{Tr}(E_{i}^{~j})
\mbox{Tr}(E_{k}^{~l})=P_{ik}^{~jl}-\frac{1}{N}I_{ik}^{~jl}.$$
So for the invariant $W^0$ one can write
$$
W^0=\beta_1\left(P-\frac{1}{N}I\right).
$$

Before considering the invariant tensors in higher dimensions let us
note that the trace of any matrix $E_{i}^{~j}\in {\cal G}$ is equal to zero
and therefore there is no Ad-invariant tensors in $\cal G$ (${\cal G}$ is
simple). It means that in $S^2{\cal G}\otimes\wedge^2{\cal G}$
there is only one invariant tensor $L$ which is obtained from
$\mbox{Tr}(E_{i}^{~j}E_{k}^{~l}E_{m}^{~n}E_{s}^{~p})$ by
symmetrization with respect
to the first pair of indexes and
antysimmetrization  with respect to the second one:
$$
L_{ik~mn}^{~jl~sp}=
\mbox{Tr}\left(E_{i}^{~j}E_{k}^{~l}E_{m}^{~s}E_{n}^{~p}\right)+
\mbox{Tr}\left(E_{k}^{~l}E_{i}^{~j}E_{m}^{~s}E_{n}^{~p}\right)-
\mbox{Tr}\left(E_{i}^{~j}E_{k}^{~l}E_{n}^{~p}E_{m}^{~s}\right)-
$$
$$
\mbox{Tr}\left(E_{k}^{~l}E_{i}^{~j}E_{n}^{~p}E_{m}^{~s}\right).
$$
Using the definition of the basis $\{E_{i}^{~j}\}$ and performing the
explicit calculations we find
$$
L_{ik~mn}^{~jl~sp}=\delta_{k}^{~j}\delta_{i}^{~s}\delta_{m}^{~p}
\delta_{n}^{~l}
+\delta_{i}^{~l}\delta_{k}^{~s}\delta_{m}^{~p}\delta_{n}^{~j}
-\delta_{i}^{~p}\delta_{k}^{~j}\delta_{n}^{~s}\delta_{m}^{~l}
-\delta_{i}^{~l}\delta_{m}^{~j}\delta_{k}^{~p}\delta_{s}^{~n}
$$
$$
-\frac{2}{N}\delta_{k}^{~l}\delta_{i}^{~s}\delta_{m}^{~p}\delta_{n}^{~j}
-\frac{2}{N}\delta_{i}^{~j}\delta_{k}^{~s}\delta_{m}^{~p}\delta_{n}^{~l}
+\frac{2}{N}\delta_{i}^{~p}\delta_{n}^{~s}\delta_{k}^{~l}\delta_{m}^{~j}
+\frac{2}{N}\delta_{i}^{~j}\delta_{m}^{~l}\delta_{k}^{~p}\delta_{n}^{~s}.
$$
Thus, the tensor $W^2\in S^2{\cal G}\otimes \wedge^2{\cal G}$ coincides
with $L$ up to the numerical constant which we call $\beta_2$:
$W^2=\beta_2L$.

Now substituting $W^0$ and $W^2$ in (\ref{d13}) we obtain the expression
for the $SL(N)$ bicovariant bracket:
\begin{equation}
\{\theta_1,\theta_2\}_{\mbox{c.b.}}=\beta_1\left(P-\frac{1}{N}I\right)+
\left\{\theta_1[\theta_2, r]\right\}
+\beta_2\left(\theta_1P\theta_2+\theta_2P\theta_1-\frac{2}{N}\theta_1\theta_1-
\frac{2}{N}\theta_2\theta_2\right).
\label{14}
\end{equation}
Comparing these brackets with the $GL(N)$-covariant brackets (\ref{20})
we see that they are obtained from (\ref{20}) under special
values of  parameters $\beta_1,\beta_3$. Thus, the Jacobi identity
for the quadratic part of eq.(\ref{14}) is satisfied as soon as
$\beta_2$ is fixed to be $\pm\sqrt{\alpha}$.
One can easily realize that the quadratic part of the brackets (\ref{14})
reproduce the covariant bracket for two odd generators on $SL(N)$
derived in \cite{AM}.

\section{Concluding remarks}
In this note we have derived the general expression for
the bicovariant brackets for odd generators of the external algebra on
a Lie group. We applied this formula to $GL(N)$ and $SL(N)$ and
reproduced the Poisson-Lie brackets that was found before.
It is interesting to note
that previously these quadratic brackets were obtained as
the solution of the total Jacobi identity involving the complete set of
generators of ${\cal M}$. Now we see that the quadratic part
of these brackets is completely fixed without any reference
to the brackets of the first order.

The proposed general expression opens the way to resolve the question if
there exist bicovariant Poisson brackets for other classical Lie groups.
Namely, one has to classify the Ad-invariant constant tensors on a
given group. As soon as these tensors are described the only task is
to impose the Jacobi identity that as we suggest is also a problem
of the transparent geometric nature.

The quantization of these brackets would lead to the graded noncommutative
algebra that can be treated as the algebra of quantum right-invariant
forms on the corresponding quantum group.
The interesting question if it is possible to supply this algebra by
the quantum exterior derivative $d$.
As it was shown in \cite{FP}
even for $SL_q(N)$ the quantum operator $d$
may be defined but it looses some usual
properties. Since
a bicovariant bracket on any Lie group contains the term
$\{\theta_1[\theta_2,r]\}$ the quantization of this part leads to the
appearance of the quantum $R$-matrix in the defining relations of the
quantum algebra. As for the other terms in the bracket defined via the
Ad-invariant tensors their quantization means that we construct the
tensors that would be the invariants of the adjoint action of the
corresponding quantum group.

$$~$$
{\bf ACKNOWLEDGMENT}
$$~$$
The authors are grateful to I.Ya.Aref'eva, I.V.Volovich, A.P.Isaev
and P.N.Pyatov for interesting discussions.

\end{document}